\documentclass[journal]{IEEEtran}
 \usepackage{amsmath,amssymb}
 \usepackage{subfigure}
 \usepackage{graphicx,graphics,color,psfrag}
 \usepackage{cite,balance}
 \usepackage{caption}
 \allowdisplaybreaks
 \usepackage{algorithm}
 \usepackage{accents}
 \usepackage{amsthm}
 \usepackage{bm}
 \usepackage{algorithmic}
 \usepackage[english]{babel}
 \usepackage{multirow}
 \usepackage{enumerate}
 \usepackage{cases}
 \usepackage{stfloats}
 \usepackage{dsfont}
 \usepackage{color,soul}
 \usepackage{amsfonts}
 \usepackage{cite,graphicx,amsmath,amssymb}
 \usepackage{subfigure}
 \usepackage{fancyhdr}
 \usepackage{hhline}
 \usepackage{graphicx,graphics}
 \usepackage{array,color}
 \usepackage{amsmath}
 \usepackage{booktabs}
\include{header}

\begin{document}
\setlength{\topskip}{-3pt}

\graphicspath{{Figs/}}

\title{6G Non-Terrestrial Networks Enabled Low-Altitude Economy: Opportunities and Challenges}
\author{Yihang Jiang, Xiaoyang Li, Guangxu Zhu, Hang Li, Jing Deng, Kaifeng Han, Chao Shen, Qingjiang Shi, and Rui Zhang
\thanks{Yihang Jiang, Xiaoyang Li, Guangxu Zhu, Hang Li, Chao Shen, and Rui Zhang are with the Shenzhen Research Institute of Big Data (SRIBD), The Chinese University of Hong Kong-Shenzhen, Guangdong, China. Jing Deng is with Wireless Center of China Mobile, Jiangxi, China. Kaifeng Han is with China Academy of Information and Communication Technology, Beijing, China. Qingjiang Shi is with Tongji University, Shanghai, China, and also with SRIBD. Corresponding author: Xiaoyang Li (e-mail: lixiaoyang@sribd.cn).}}


\maketitle

\begin{abstract}
The unprecedented development of \emph{non-terrestrial networks} (NTN) utilizes the low-altitude airspace for commercial and social flying activities. The integration of NTN and terrestrial networks leads to the emergence of \emph{low-altitude economy} (LAE). A variety of LAE application scenarios are enabled by the sensing, communication, and transportation functionalities of the aircrafts. The technological prerequisites supporting LAE are introduced in this paper, including three-dimensional network coverage and aircrafts detection. Then, the aircraft-assisted sensing and communication functionalities essential to LAE are reviewed, including the terrestrial and non-terrestrial targets sensing, ubiquitous coverage, relaying, and traffic offloading. Finally, several future directions are identified, including aircrafts collaboration, energy efficiency, and artificial intelligence enabled LAE. 
\end{abstract}

\section{introduction}
As a new economic form, \emph{low-altitude economy} (LAE) utilizes low-altitude airspace (generally referring to the space within 1000 meters above the ground) to carry out various flying activities, creating commercial and social values. The flying activities are conducted by various manned and unmanned aircrafts in \emph{non-terrestrial networks} (NTN), enabling a variety of applications including \emph{internet of things} (IoT), \emph{artificial intelligence} (AI), transportation, logistics, tourism, agriculture, and disaster monitoring \cite{giordani2020non}. 

The concept of LAE has attracted world-wide attentions. In the United States, the Federal Aviation Administration has announced the construction standards for LAE, and has led in-depth cooperation with the National Aeronautics and Space Administration to provide intelligent and efficient low-altitude management services. In Canada, the Transport Canada has established regulations and guidelines to ensure the safe and responsible operations in LAE. The European Union has implemented policies for LAE under the regulation of the European Union Aviation Safety Agency. In the United Kingdom, the rules and guidelines for LAE have been set by the Civil Aviation Authority. In China, a series of policies for supporting LAE have been published by the Civil Aviation Administration of China. Multiple companies have participated in LAE, including Da-Jiang Innovations and Parrot for manufacturing aircrafts, Amazon Prime, Wing, Meituan, and Shunfeng Express for logistics and delivery, Skydio and Autel Robotics for aerial photography and filmmaking, as well as PrecisionHawk and Kespry for inspection and surveying.

\newpage However, the deployment of LAE still faces several practical challenges. According to the Statista, the number of aircrafts will reach 9.6 million in 2030 \cite{gupta2015survey}. The increasing number of aircrafts makes it hard to design obstacle avoidance systems for aircrafts \cite{meng2023uav}. Moreover, the information exchange between multiple aircrafts and their control centers results in heavy burden on communication overheads \cite{fei2023air}. To deal with the above issues, the aircrafts need to be surveilled and regulated \cite{cui2023toward}. However, the existing aircrafts mainly rely on simple point-to-point communication over the unlicensed band, which renders them difficulty for surveillance and regulation. The development of the \emph{sixth-generation} (6G) mobile network provides potential solutions to tackle this issue \cite{mozaffari2021toward}. Moreover, the spectrum sharing technology enables the opportunistic spectrum access of aircrafts for communication. The \emph{integrated sensing and communication} (ISAC) technology further supports simultaneous surveillance and message passing of aircrafts \cite{mu2023uav}.

To meet the need for seamless communication coverage, 6G will no longer be limited to terrestrial networks. The integration of aircrafts and terrestrial networks leads to the \emph{integrated air-ground network} (IAGN). As shown in Fig. 1, the components of IAGN include terrestrial and aerial \emph{base stations} (BSs) and \emph{user equipments} (UEs), as well as a number of infrastructures including take-off and landing platforms, power-charging stations, and logistics distribution centers to support the functionalities of sensing, communication, and transportation in LAE. A digital management service system needs to be developed to provide the services including flight planning, navigation, and surveillance for aircrafts. Several typical working paradigms involved are discussed below: 

\begin{itemize}
\item \emph{Control and non-payload communication (CNPC):} As a bidirectional communication link, CNPC plays a key role in IAGN to ensure the control and navigation of aircrafts. CNPC typically operates at low data rates but demands exceptionally high levels of reliability, stringent security measures, and minimal latency for continuous connectivity. Specifically, for the non/semi-autonomous aircrafts, the control station needs to transmit control commands in real time or periodically to guide the flights of the aircrafts. In turn, the aircrafts need to report their flight status (such as the flight altitude and velocity) as well as position information via the uplink CNPC so that the control station can safely regulate their flights. 

\item \emph{Payload Communication (PC):} PC refers to mission-related communications depending on the application scenarios. Compared with CNPC, PC usually has higher tolerance of latency and security requirements. The PC links of aircrafts could reuse the existing spectrum such as the long-term evolution band for cellular coverage, or use new spectrum such as millimeter wave band for high-capacity wireless backhaul.

\item \emph{Sensing:} In addition to communications, the robust sensing capability of IAGN is another key technology to support LAE. Intuitively, the sensing can be performed based on the radar reflection signals. Besides radio frequency sensing, the on-board sensors of the aircraft also provide multimodal sensing information. These technologies provide accurate environmental information and also greatly assist LAE in its various business activities. 
\end{itemize}

In this article, we provide an overview of LAE by illustrating its main application scenarios in Section II, introducing its technological prerequisites in Section III, delving into its unique aircraft-assisted sensing and communication functionalities for LAE in Section IV, and discussing its future directions in Section V. To the best of the authors’ knowledge, this article is the first to present a comprehensive overview of LAE and its enabling 6G NTN technologies.

\begin{figure*}[!t]
  \centering
  \includegraphics[scale = 0.55]{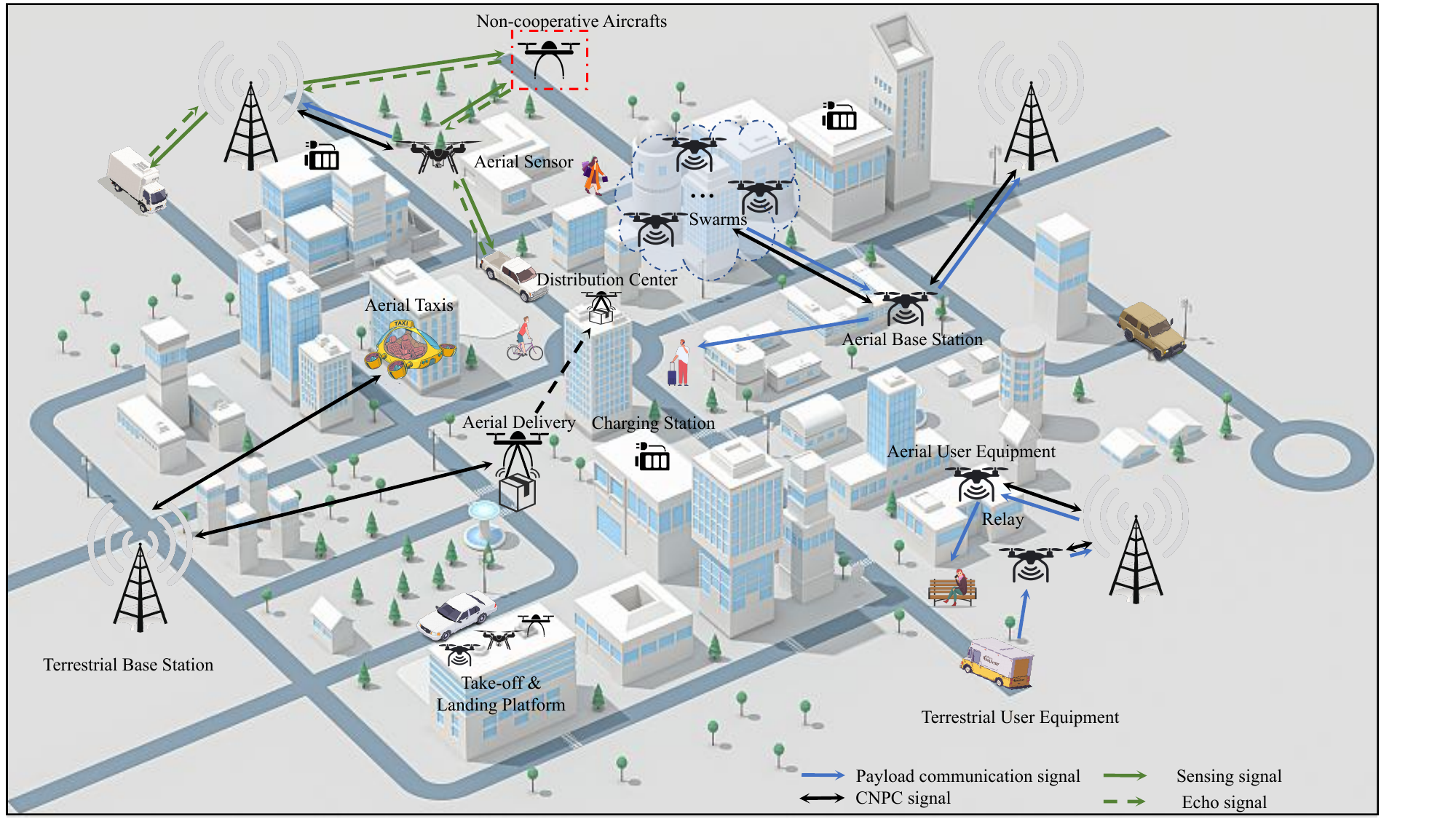}
  \caption{IAGN network architecture for supporting LAE.}
  \label{FigIAGN}
\end{figure*}

\section{Application Scenarios of LAE}
\begin{figure*}[!t]
  \centering
  \includegraphics[scale = 0.5]{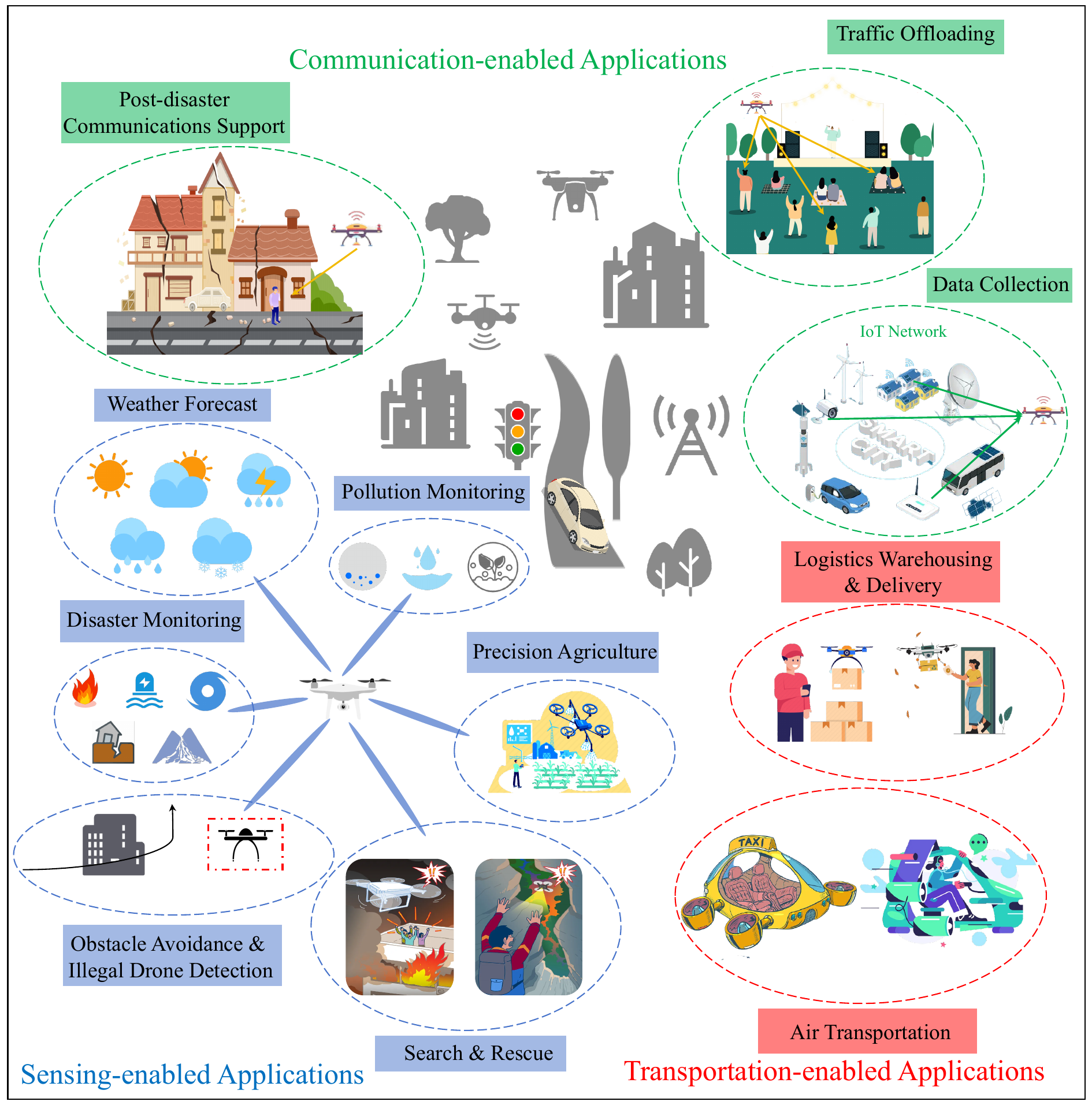}
  \caption{Application scenarios of LAE.}
  \label{FigApp}
\end{figure*}

As shown in Fig.~\ref{FigApp}, the application scenarios of LAE are generally enabled by three functionalities of the aircrafts, namely communication, sensing, and transportation.

\subsection{Communication-Enabled Application Scenarios}
The communication capabilities of aircrafts enable a series of applications \cite{zeng2019accessing}. One of the most prominent application scenarios is adopting aircrafts as \emph{aerial BSs} (ABSs) to achieve ubiquitous coverage. The ABSs are deployed in remote areas without cellular infrastructure or crowded areas such as sports events with unusually large user demand. The ABSs also play a critical role in supporting the post-disaster communication service restoration. Moreover, the aircrafts can be deployed as relays connecting BS and UE without reliable direct communication links. The aircrafts can also be used for information dissemination and collection of delay-tolerant messages to/from the distributed wireless devices.

\subsection{Sensing-Enabled Application Scenarios}
In addition to the communication-enabled application scenarios, the sensing capabilities of aircrafts can also be exploited to support numerous applications \cite{toro2018uav}. First, sensing plays a key role in obstacle avoidance, which guarantees the safety and reliability of the flight. Second, in terms of airspace management, legal aircrafts can work with BS to detect the illegal drones and safeguard them if necessary. Third, the multimodal sensors equipped on aircrafts can collect information about temperature, humidity, and atmospheric pressure, which can be applied for weather forecasting. Aircrafts equipped with high-resolution cameras have revolutionized the field of aerial photography and videography, which can offer unique perspectives and capture stunning visuals for various purposes like filmmaking, real estate marketing, event coverage, and tourism promotion. Aircrafts equipped with advanced sensors provide efficient and accurate data for surveying and mapping purposes, which assist in creating digital elevation models, \emph{three-dimensional} (3D) maps, and topographic surveys for urban planning, construction, and land management. The aircrafts can also be used for monitoring environment changes including air quality, water quality, and land pollution. In addition, aircrafts can be used to monitor various natural disasters, such as early warning and post-disaster assessment of fires, floods, typhoons, earthquakes and landslides. In agriculture, crop monitoring and assessment, as well as precise pest control, fertilization and spraying, can be achieved using the information sensed by the aircrafts. With flexible maneuverability, the aircrafts can penetrate into hard-to-reach areas executing more efficient search and rescue missions.

\subsection{Transportation-Enabled Application Scenarios}
Beyond on-board facilities, aircrafts fundamentally serve as carriers with flight capabilities \cite{menouar2017uav}. Rapid and convenient goods delivery services for various scenarios are available with the efficient aerial transportation capabilities of aircrafts. In logistics warehousing and supply chain, aircrafts can be used for cargo dispatching to improve logistics efficiency. In urban settings, aircrafts can be employed for swift delivery of goods such as parcels and food orders, offering an efficient “last-mile” transportation solution. In remote or hard-to-reach areas, aircrafts can also replace manual mail/parcel delivery. Moreover, the aircrafts can provide short-distance transportation services for passengers in low amplitude space, which are known as aerial taxis. The manned aircraft has already been employed for emergency medical transportation, which greatly improves the efficiency of rescues.


\section{Technological Prerequisites for LAE}
\begin{figure*}[t]
  \centering
  \includegraphics[scale = 0.54]{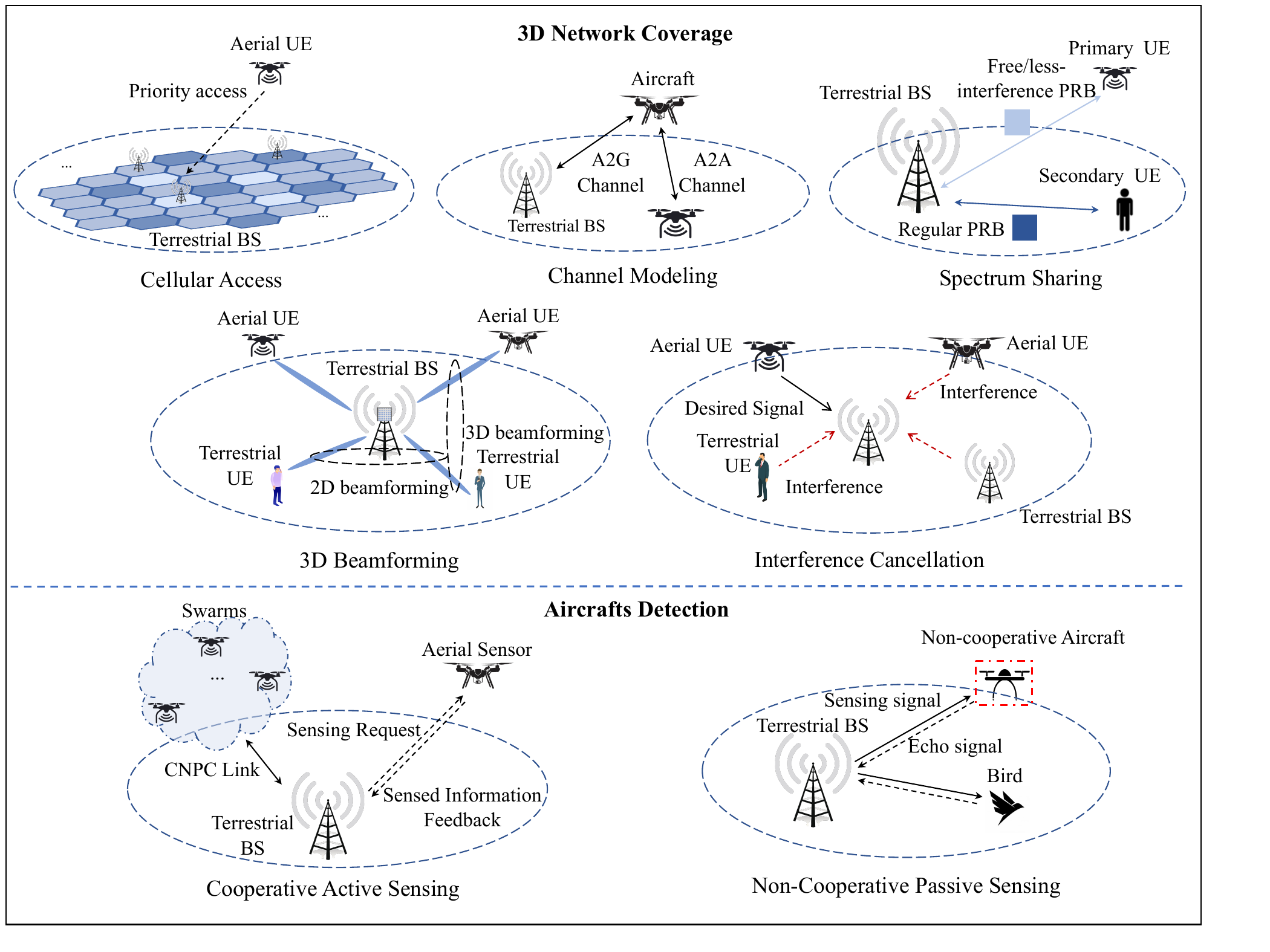}
  \caption{Technological prerequisites supporting LAE.}
  \label{FigChallenge}
\end{figure*}

In this section, the technological prerequisites supporting LAE including the 3D network coverage and aircrafts detection as depicted in Fig.~\ref{FigChallenge} are discussed as follows.

\subsection{3D Network Coverage}
To improve the communication performance, the aircrafts are expected to be connected to the cellular networks \cite{zeng2018cellular}. Cellular-connected aircraft is an appealing solution in practical implementation, which only needs to reuse the existing cellular architecture and facilities. However, the performance of the existing terrestrial users in cellular network will be affected. A series of efforts need to be made to resolve the 3D coverage problem for low-altitude aircrafts. 

\subsubsection{Cellular Access}
For initial cellular access, BSs regularly transmit \emph{synchronization signal blocks} (SSBs) that facilitate the cell search and selection for aircrafts. Based on the received SSBs, the aircrafts select the best cell to access, where the channel conditions between the best cell and the aircrafts has the strongest reference signal received power. Note that unlike traditional terrestrial cellular communications, the strong air-to-ground \emph{line-of-sight} (LoS) channels allow aircrafts to connect more BSs, which makes the aircrafts less inclined to access their physically nearest BS. Moreover, as the high mobility of aircrafts would introduce frequent cell switch, the multi-cell cooperation may bring larger macro-diversity gain than conventional single-cell access. Since the multi-cell cooperation will inevitably entail more resource overhead for information exchange as well as higher computational complexity for joint processing, further investigation is needed for evaluating its practical performance.

\subsubsection{Channel Modeling}
Due to the aircraft's altitude and high mobility, aircraft communications (both \emph{air-to-ground} (A2G) and \emph{air-to-air} (A2A)) have distinguish channel characteristics from the conventional \emph{ground-to-ground} (G2G) communications. It is characterized by highly dynamic communication channel characteristics and excessive spatial and temporal variations. In addition to buildings and terrain, the A2G communication channels are also affected by the airframe shadowing during aircraft flight. The A2A communication channels are mainly dominated by the LoS component and the effect of multi-path is negligible compared to A2G or G2G communication channels. Moreover, the high speed of aircrafts leads to excessive Doppler frequency shifts. This will pose a great challenge on the modeling and measurements of the A2A channels. To improve the performances of A2G and A2A communications in LAE, it is importance to exactly characterize the channel models. As a vision for 6G, digital twins could be a potential solution. For example, a digital airspace map of the low altitude containing information about the low altitude environment and other information related to low altitude flight could be utilized to support channel modeling.

\begin{figure*}[t]
  \centering
  \includegraphics[scale = 0.54]{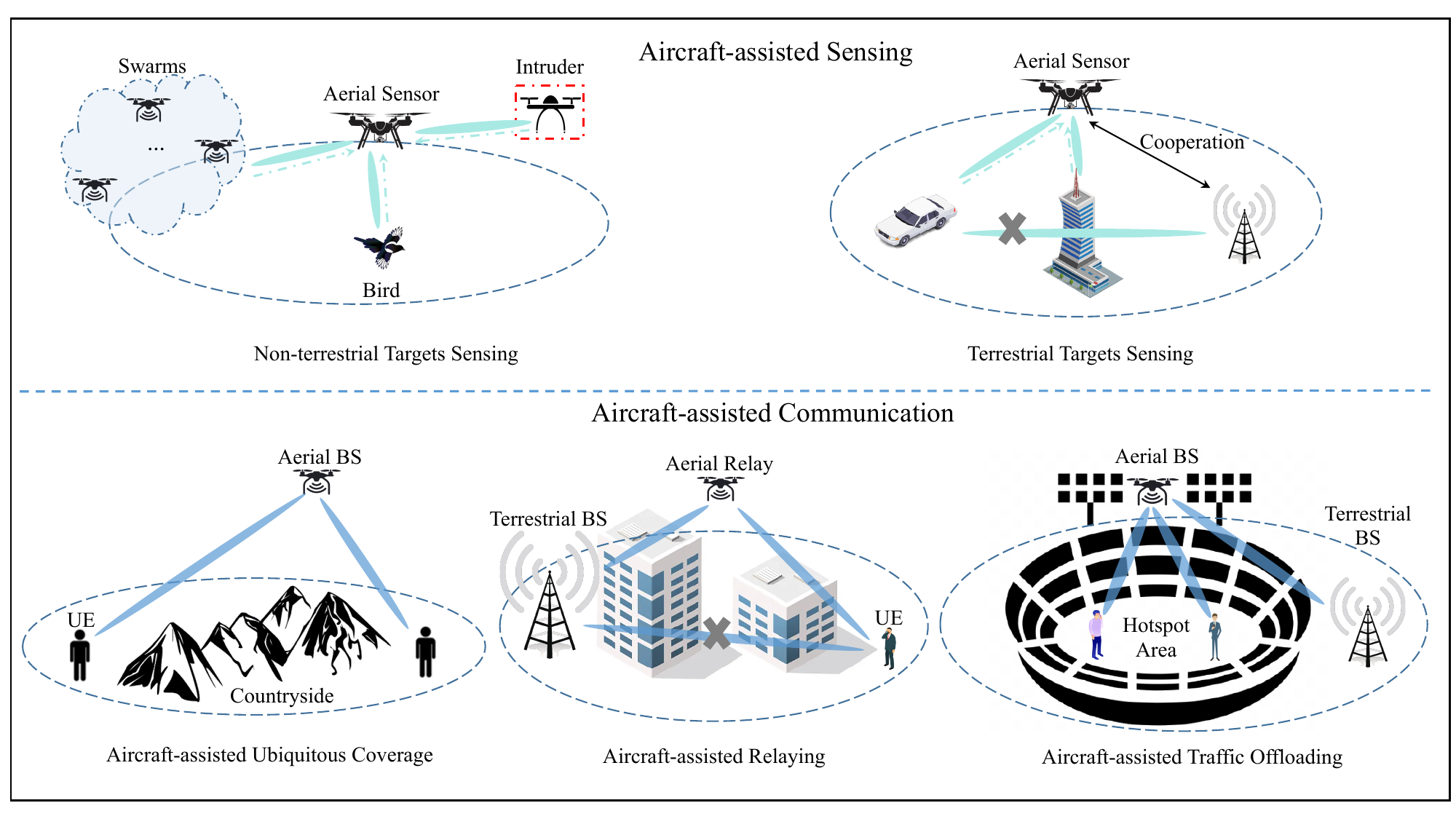}
  \caption{Aircraft-assisted sensing and communication.}
  \label{FigAirAssist}
\end{figure*}

\subsubsection{Spectrum Sharing}
The spectrum sensing capabilities of aircrafts can also be utilized to sense the network conditions and ensure efficient spectrum allocation for communication. By defining the priorities of different UEs/tasks, cognitive radio can allocate free/less-interfering \emph{physical resource blocks} (PRBs) to different UEs/tasks via spectrum sensing methods such that the reliable connections with less interference can be achieved. Existing classical spectrum sensing methods include energy detection, matched filter detection, and periodicity detection. Combining these existing spectrum sensing methods and based on the network data collected by the aircrafts, effective PRBs allocation can be realized through the collaboration between the aircrafts and the BSs to reduce the spectrum burden. 

\subsubsection{3D Beamforming}
The BSs in existing cellular networks are typically equipped with a full-dimensional large-scale antenna array, which enables the fine-grained 3D beamforming with a high degree of configurability. Unlike the traditional 2D beamforming with fan-shaped beams for terrestrial communications, 3D beamforming can better mitigate the interference among high-altitude aircrafts and terrestrial UEs with the fine-grained beams in both the azimuth and elevation dimensions, and thus leads to better communication performance. Note that the performance of most beamforming techniques depends heavily on accurate channel state information. Therefore, accurate channel/beam tracking techniques are required due to frequent channel variations arising from the high mobility of aircraft.

\subsubsection{Interference Cancellation}
Compared to terrestrial equipment, the relatively high altitude of aircrafts results in more dominant LoS links and thus wider coverage. However, the LoS links also introduce severe interference. In terrestrial cellular networks, adjacent BSs can dynamically allocate PRBs to their respective UEs based on the shared control information, so as to avoid the inter-cell interference. However, for cellular-connected A2G an A2A communications that are dominated by LoS paths, there are a much larger number of BSs with effective co-channel signals. The control information exchange between these BSs can be difficult or even infeasible at the same time due to the severe co-channel interference. Therefore, interference cancellation deserves further research in order to guarantee the performance of cellular-connected aircrafts communication in LAE.

\subsection{Aircrafts Detection}
Depending on whether the aircrafts participate in the sensing process or not, the technologies for aircrafts detection can be generally divided into two categories, namely cooperative active sensing and non-cooperative passive sensing.

\subsubsection{Cooperative Active Sensing}
Conventional sensing relies on the aircraft's on-board sensors, including video cameras for visual sensing and environment recognition, inertial measurement units for estimating acceleration and angular velocity, global navigation satellite system for localization, and radar for detection. Reliable surveillance can be achieved by making full use of such multimodal sensing information. However, considering the limited power supply of the aircrafts in practice, the multimodal sensing information is not easy to obtain. An alternative surveillance scheme is to actively sense the aircrafts from the BS through cooperation. Specifically, a sensing request is initiated by the BS and accepted by the aircrafts, the sensing information is fed back to the BS, which can be data from multimodal sensors and/or channel estimation based on the pilot signal. Based on the feedback, the BS can better direct the aircrafts' flight from a global perspective.

\subsubsection{Non-Cooperative Passive Sensing}
Non-cooperative passive sensing is mainly used to detect the illegal aircrafts. The activities of illegal aircrafts result in obstructing the movement and interfering the communication of legal aircraft. To resolve this issue, radar sensing based on echoed/scattered signals can be utilized to detect, classify and track illegal aircrafts to ensure the safety of the legal aircraft. However, low-altitude aircrafts are relatively smaller in size and thus have lower radar cross sections compared to conventional high-altitude aircrafts, which poses a great challenge for detection and tracking. In addition, their slower flight speeds and hovering capabilities make it difficult to distinguish them from the static clutters and birds. Accounting for this problem, the micro-Doppler signatures of small aircrafts' movements need to be captured to classify the small aircrafts and birds.

\begin{figure*}[t]
  \centering
  \subfigure[Aircraft-assisted ISAC system \label{FigSys}]
      {\includegraphics[width=0.4\linewidth]{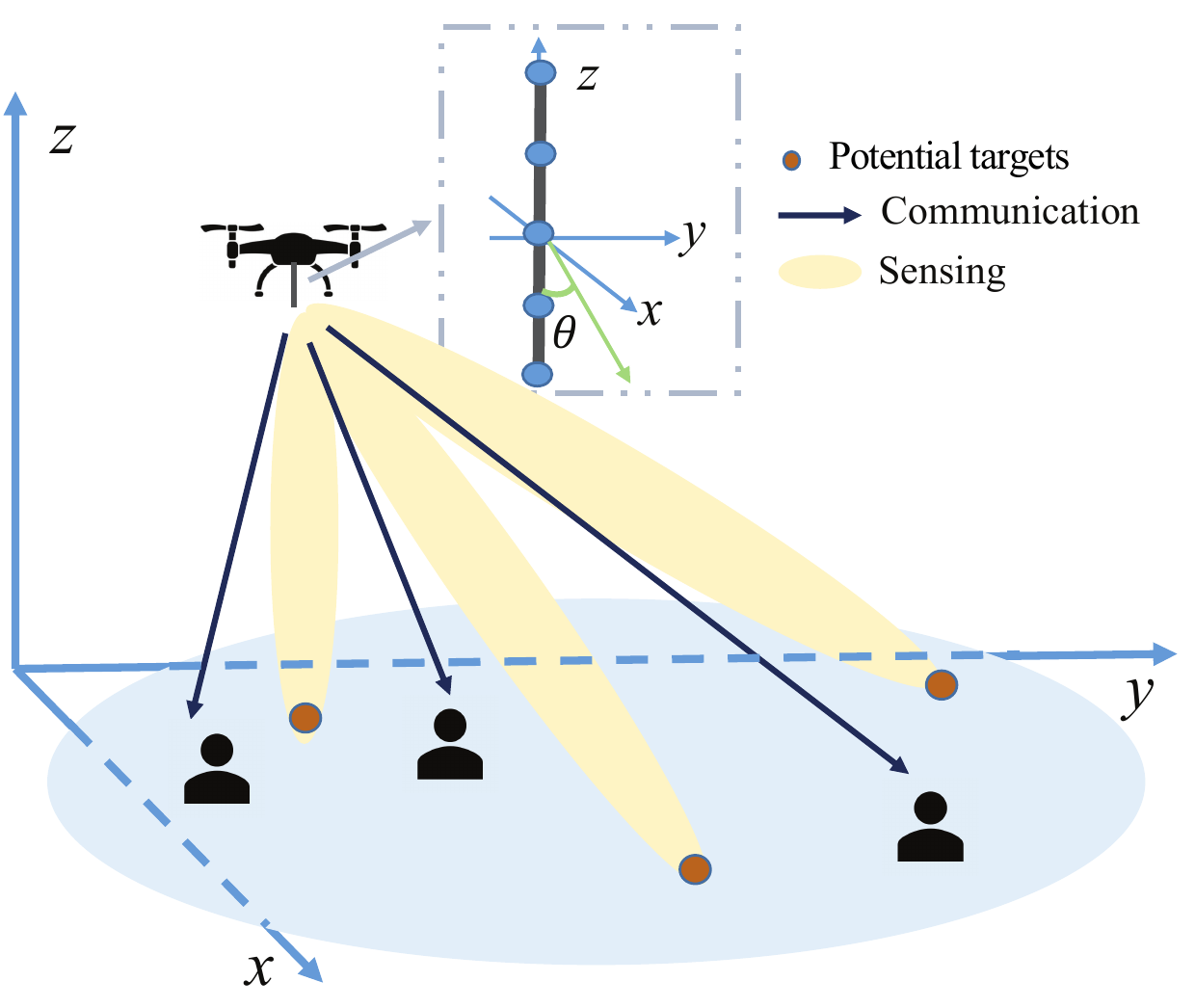}}
  \subfigure[Optimized trajectories under different design objectives\label{FigSim}]
      {\includegraphics[width=0.5\linewidth]{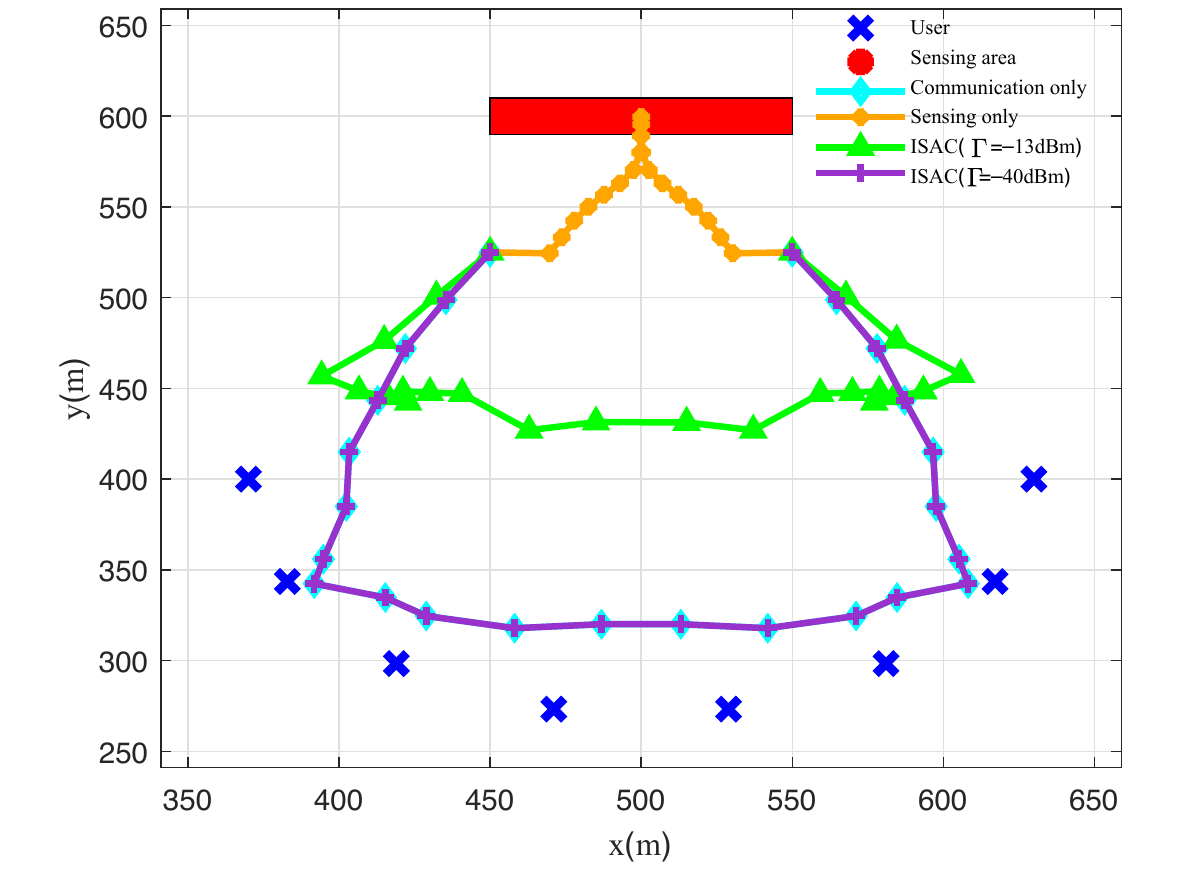}}  
  \caption{Aircraft-assisted ISAC.} 
  \label{FigISAC}
  \end{figure*}
  
\section{Aircraft-assisted Functionalities in LAE}
In this section, the functionalities assisted by aircrafts in LAE as depicted in Fig.~\ref{FigAirAssist} are reviewed.

\subsection{Aircraft-assisted Sensing}
Benefiting from the high mobility and flexibility, the aircrafts can provide more degrees of freedom to support sensing. Different from the terrestrial sensors with detection blind area, the aircrafts are expected to achieve 360-degree sensing. The aircraft-assisted sensing for terrestrial and non-terrestrial targets are discussed in the following.

\subsubsection{Non-terrestrial Target Sensing}
Non-terrestrial targets sensing mainly refers to the detection of other aircrafts and birds. For a typical aircraft, other aircrafts can be cooperators in a specific task or non-cooperative intruders. The cooperative aircrafts form a swarm, in which aircrafts can detect each other and establish a reliable coordination for safe flight even when the swarm loses the remote control. The distances and angles of the non-cooperative intruders can also be detected by the aircrafts to avoid collision. Accounting for the mobility of non-terrestrial targets, beam tracking is needed to perceive the desired target in real time. However, the flight stability of the aircraft is affected by the bumpy air flow, which increases the difficulty of beam tracking.

\subsubsection{Terrestrial Targets Sensing}
The aircrafts can also be deployed for sensing terrestrial targets in a cooperative manner. By sharing the sensed information of different aircrafts, larger sensing coverage and more accurate target parameter estimation can be achieved. Moreover, the multi-modal sensors equipped on aircrafts can collect different types of target information. The integration of these features can further improve the sensing performance. However, it should be noted that only the LoS links of aircrafts are exploited for sensing, while the NLoS links are treated as unfavorable interference. To mitigate the interference, the beamforming design, power control, and trajectory planning of aircrafts need to be investigated.

\subsection{Aircraft-assisted Communication}
The flexibility and mobility of aircrafts can be exploited to support different communication tasks. The typical application scenarios are discussed below.
  
\subsubsection{Aircraft-assisted Ubiquitous Coverage}
For areas where basic communication facilities are not available, aircrafts can be used as ABSs to enhance the coverage and performance of communication networks. The ABS coverage problem was first considered in \cite{al2014optimal}, where the channel models between ABSs and UEs were established to determine the ABS's optimal altitude to maximize the coverage area. The deployment of ABSs still faces several challenges in practice. First, the horizontal/vertical placement of multiple ABSs should be investigated to cover UEs in wider areas. Second, the endurance of aerial BSs is constrained by the limited battery equipped on aircrafts, which needs to be considered for ABSs deployment design. Moreover, when multiple ABSs are deployed, the effect of their potential interference should be considered.

\subsubsection{Aircraft-assisted Relaying}
In the absence of reliable communication links between BSs and UEs, aircrafts can serve as the relays. With LoS-dominant links, aircrafts can capture, amplify, and relay communication signals to target locations, thereby improving wireless connectivity quality in long-distance communication scenarios. It is worth noting that, unlike ABSs, utilizing aircrafts as aerial relays can reduce the burden on the on-board equipment. For the network consisting of multiple aerial relays, the optimal assignment of UAV relays to different tasks to maximize the overall utility needs to be investigated in future work.

\subsubsection{Aircraft-assisted Traffic Offloading}
In conventional terrestrial cellular systems, it is difficult for the BSs to provide effective support for UEs at cell edges or hotspot areas. By exploiting the mobility and enhanced communication capability of LoS links, aircrafts-assisted cellular offloading provides a promising solution. It is worth noting that hotspots tend to have higher UE densities compared to non-hotspots. To guarantee the communication performance in hotspot areas, the flight trajectory of aircraft and radio resource allocation need to be jointly designed, given the constraints on the total system bandwidth and aircraft transmit power.

\subsection{Aircraft-assisted ISAC}
To improve the spectrum efficiency, the aircrafts can transmit ISAC signals to communicate with multiple UEs and sense targets simultaneously. Aiming at improving the communication performance while guaranteeing the sensing performance, the flight trajectories and beamforming of the aircrafts were jointly designed in \cite{lyu2022joint}. The flight trajectories of aircrafts under different designs are plotted in Fig.~\ref{FigISAC}. It can be observed that if only sensing is considered, the aircraft directly flies to the sensing target. If only communication is considered, the aircraft approaches the UEs as closer as possible under realistic flight constraints. As for the ISAC case, the aircraft moves between the sensing target and UEs. When the sensing beampattern gain threshold is relatively large (--13 dBm), the flight trajectory of the aircraft is closer to the sensing target. In contrast, when the sensing beampattern gain threshold is relatively small (--40 dBm), the flight trajectory of the aircraft is closer to the UEs.

\begin{figure*}[t]
  \centering
  \includegraphics[scale = 0.54]{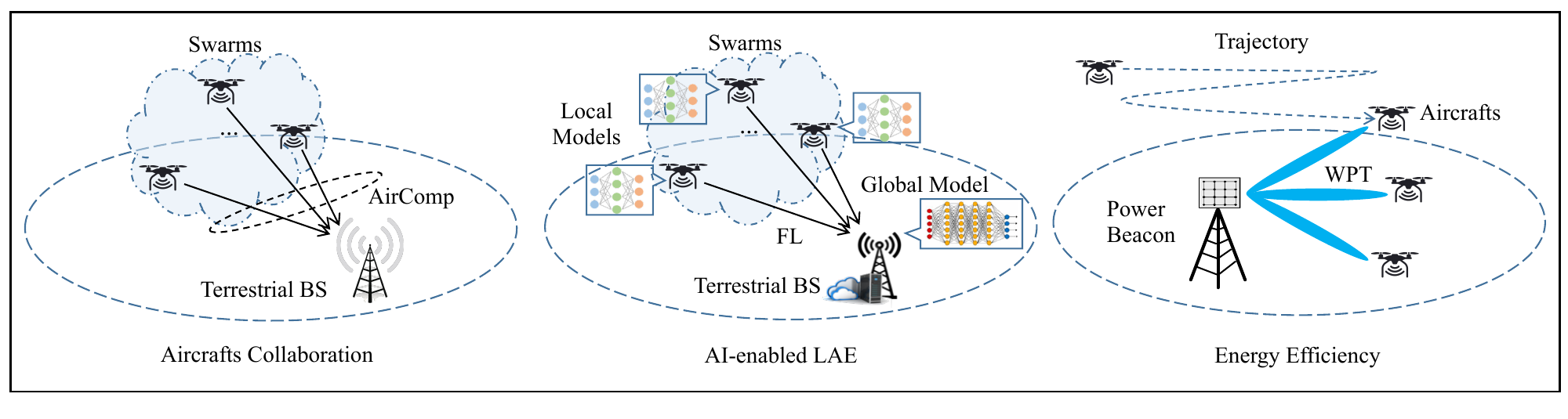}
  \caption{Future directions of LAE.}
  \label{FigFuture}
\end{figure*}

\section{Future Directions of LAE}
As depicted in Fig.~\ref{FigFuture}, there are multiple future directions in LAE worthy of further investigation.

\subsection{Aircrafts Collaboration}
In LAE, multiple aircrafts need to share their sensed information with BS for collaboration. The messages delivered by different aircrafts will compete for the radio resource in conventional multi-access schemes, which will result in transmission latency that is intolerable for delay-sensitive tasks such as collision avoidance. To facilitate the information sharing process, \emph{over-the-air computation} (AirComp) has been proposed that utilizes the waveform superposition property of wireless signals to aggregate the data simultaneously transmitted by multiple aircrafts \cite{li2023integrated}. However, the mobility of the aircraft makes it harder to balance the channels for guaranteeing the computation accuracy. Therefore, the information sharing process for aircrafts collaboration in LAE deserves further study.

\subsection{AI-enabled LAE}
The development of AI is expected to support more intelligent tasks in LAE, such as smart logistics and auto-driving. The AI models are trained based on the data sensed by the aircrafts. However, the limited computation capabilities of aircrafts might fail to support the training process of sophisticated AI models, while offloading the raw data to central server for computation will cause privacy leakage. To deal with this problem, \emph{federated learning} (FL) enables each aircraft to update its local model based on the sensed data and send the local updated results to the central server for the global model update \cite{tang2023integrated}. In LAE, the parameters relevant to aircrafts such as trajectory and velocity might have significant effects on FL performance, which warrants further investigation.

\subsection{Energy Efficiency}
Aircrafts are often energy-constrained devices and therefore energy allocation for different tasks needs to be considered. For example, transmit power and flight trajectories need to be optimized to carry out activities in an energy-efficient manner, thereby reducing the frequency of battery charging/replacement. Also, the location of charging stations can also be optimized to improve the efficiency of charging the vehicles in the IAGN such as UAVs and air taxis. Notably, \emph{wireless power transfer} (WPT) has been widely used in various systems to power the devices in short distances. The application of WPT in LAE is a potential solution to enable automatic charging of aircrafts, while its impact on energy efficiency needs to be further investigated.

\section{Conclusion}
In the future 6G wireless networks, NTN is expected to be integrated with terrestrial networks to support LAE. The LAE application scenarios enabled by the sensing, communication, and transportation functionalities of the aircrafts were introduced in this paper. The technological prerequisites supporting LAE were also discussed, including 3D network coverage and aircrafts detection. Furthermore, an overview of sensing and communication functionalities assisted by aircrafts was provided, including the non-terrestrial and terrestrial targets sensing, ubiquitous coverage, relaying, and traffic offloading. Finally, promising future directions worthy of investigation were highlighted, such as aircrafts collaboration, AI-enabled LAE, and energy efficiency. 

\bibliographystyle{IEEEtran}
\bibliography{refs}

\begin{thebibliography}{10}
\providecommand{\url}[1]{#1}
\csname url@samestyle\endcsname
\providecommand{\newblock}{\relax}
\providecommand{\bibinfo}[2]{#2}
\providecommand{\BIBentrySTDinterwordspacing}{\spaceskip=0pt\relax}
\providecommand{\BIBentryALTinterwordstretchfactor}{4}
\providecommand{\BIBentryALTinterwordspacing}{\spaceskip=\fontdimen2\font plus
\BIBentryALTinterwordstretchfactor\fontdimen3\font minus
  \fontdimen4\font\relax}
\providecommand{\BIBforeignlanguage}[2]{{%
\expandafter\ifx\csname l@#1\endcsname\relax
\typeout{** WARNING: IEEEtran.bst: No hyphenation pattern has been}%
\typeout{** loaded for the language `#1'. Using the pattern for}%
\typeout{** the default language instead.}%
\else
\language=\csname l@#1\endcsname
\fi
#2}}
\providecommand{\BIBdecl}{\relax}
\BIBdecl

\bibitem{giordani2020non}
M.~Giordani and M.~Zorzi, ``Non-terrestrial networks in the {6G} era:
  {Challenges} and opportunities,'' \emph{IEEE Netw.}, vol.~35, no.~2, pp.
  244--251, 2020.

\bibitem{gupta2015survey}
L.~Gupta, R.~Jain, and G.~Vaszkun, ``Survey of important issues in {UAV}
  communication networks,'' \emph{IEEE Commun. Surveys Tuts.}, vol.~18, no.~2,
  pp. 1123--1152, 2015.

\bibitem{meng2023uav}
K.~Meng, Q.~Wu, J.~Xu, W.~Chen, Z.~Feng, R.~Schober, and A.~L. Swindlehurst,
  ``{UAV}-enabled integrated sensing and communication: {Opportunities} and
  challenges,'' \emph{IEEE Wireless Commun.}, early access, 2023.

\bibitem{fei2023air}
Z.~Fei, X.~Wang, N.~Wu, J.~Huang, and J.~A. Zhang, ``Air-ground integrated
  sensing and communications: {Opportunities} and challenges,'' \emph{IEEE
  Commun. Mag.}, vol.~61, no.~5, pp. 55--61, 2023.

\bibitem{cui2023toward}
Y.~Cui, Z.~Feng, Q.~Zhang, Z.~Wei, C.~Xu, and P.~Zhang, ``Toward trusted and
  swift {UAV} communication: {ISAC}-enabled dual identity mapping,'' \emph{IEEE
  Wireless Commun.}, vol.~30, no.~1, pp. 58--66, 2023.

\bibitem{mozaffari2021toward}
M.~Mozaffari, X.~Lin, and S.~Hayes, ``Toward {6G} with connected sky: {UAVs}
  and beyond,'' \emph{IEEE Commun. Mag.}, vol.~59, no.~12, pp. 74--80, 2021.

\bibitem{mu2023uav}
J.~Mu, R.~Zhang, Y.~Cui, N.~Gao, and X.~Jing, ``{UAV} meets integrated sensing
  and communication: {Challenges} and future directions,'' \emph{IEEE Commun.
  Mag.}, early access, 2023.

\bibitem{zeng2019accessing}
Y.~Zeng, Q.~Wu, and R.~Zhang, ``Accessing from the sky: {A} tutorial on {UAV}
  communications for {5G} and beyond,'' \emph{Proc. IEEE}, vol. 107, no.~12,
  pp. 2327--2375, 2019.

\bibitem{toro2018uav}
F.~G. Toro and A.~Tsourdos, \emph{{UAV} sensors for environmental
  monitoring}.\hskip 1em plus 0.5em minus 0.4em\relax MDPI, 2018.

\bibitem{menouar2017uav}
H.~Menouar, I.~Guvenc, K.~Akkaya, A.~S. Uluagac, A.~Kadri, and A.~Tuncer,
  ``{UAV}-enabled intelligent transportation systems for the smart city:
  {Applications} and challenges,'' \emph{IEEE Commun. Mag.}, vol.~55, no.~3,
  pp. 22--28, 2017.

\bibitem{zeng2018cellular}
Y.~Zeng, J.~Lyu, and R.~Zhang, ``Cellular-connected {UAV: Potential},
  challenges, and promising technologies,'' \emph{IEEE Wireless Commun.},
  vol.~26, no.~1, pp. 120--127, 2018.

\bibitem{al2014optimal}
A.~Al-Hourani, S.~Kandeepan, and S.~Lardner, ``Optimal {LAP} altitude for
  maximum coverage,'' \emph{IEEE Wireless Commun. Lett.}, vol.~3, no.~6, pp.
  569--572, 2014.

\bibitem{lyu2022joint}
Z.~Lyu, G.~Zhu, and J.~Xu, ``Joint maneuver and beamforming design for
  {UAV}-enabled integrated sensing and communication,'' \emph{IEEE Trans.
  Wireless Commun.}, vol.~22, no.~4, pp. 2424--2440, 2022.

\bibitem{li2023integrated}
X.~Li, F.~Liu, Z.~Zhou, G.~Zhu, S.~Wang, K.~Huang, and Y.~Gong, ``Integrated
  sensing, communication, and computation over-the-air: {MIMO} beamforming
  design,'' \emph{IEEE Trans. Wireless Commun.}, vol.~22, no.~8, pp. 5383 --
  5398, 2023.

\bibitem{tang2023integrated}
Y.~Tang, G.~Zhu, W.~Xu, M.~H. Cheung, T.-M. Lok, and S.~Cui, ``Integrated
  sensing, computation, and communication for {UAV}-assisted federated edge
  learning,'' \emph{arXiv preprint arXiv:2306.02990}, 2023.

\end{thebibliography}

\end{document}